\documentclass[11pt,draft]{article}

\title{\bf The Hulth$\acute{e}$n Potential in \textsl D-dimensions.}
\author{\bf D.Agboola.\footnote{E-Mail:tomdavids2k6@yahoo.com}}
\date{Department of Pure and Applied Mathematics,\linebreak Ladoke Akintola University of Technology,Oyo State, Nigeria. \linebreak  P.M.B. 4000}

\begin{document}
\maketitle
\maketitle

\vspace{0.5in}
\noindent {\bf Abstract} An approximate solution of the Schr$\ddot{o}$dinger equation with the  Hulth$\acute{e}$n potential is obtained in \textsl D-dimensions with an exponential approximation of the centrifugal term. Solution to the corresponding hyper-radial equation is given using the conventional Nikiforov-Uvarov method. The normalization constants for the Hulth$\acute{e}$n potential are also computed. The expectation values $\langle r^{-2}\rangle$,$\langle V(r)\rangle$, are also obtained using the Feynman-Hellmann theorem.\\

\maketitle
\vspace{0.5in}
\vspace{1.2in}
\maketitle


\pagebreak
\noindent{\bf 1. Introduction.}\\\\
The search for exact bound-state solutions of wave equations, relativistic or non- relativistic, has been an important research area in quantum mechanics. However, over the past decades, problems involving the multidimensional Schr$\ddot{o}$dinger equation have been addressed by many researchers. For instance, Bateman investigated the relationship between the hydrogen atom and a harmonic oscillator potential in arbitrary dimensions [1]. The \textsl N-dimensional Kratzer-Fues potential was discussed by Oyewumi [2], while the modified Kratzer-Fues potential plus the ring shape potential in \textsl D-dimensions by the Nikiforov-Uvarov method has also been considered [3]. Very recently, some quantum mechanical properties of the pseudoharmonic oscillator were discussed by Agboola {\it et al} [4]. 

The Hulth$\acute{e}$n potential is one of the important short-range potentials which behaves like a Coulomb potential for small values of $r$ and decreases exponentially for large values of $r$. The Hulth$\acute{e}$n potential has received extensive study in both relativistic and non-relativistic quantum mechanics [5, 6, 7, 8]. Unfortunately, quantum mechanical equations with the Hulth$\acute{e}$n potential can be solved analytically only for the $s$-states [9, 10, 11]. However, a number of methods have been used to find the bound-state energies numerically [10-16] and analytically [17, 18].

Recently, an extension of this study to a multidimensional Hulth$\acute{e}$n potential in a Klein-Gordon equation was presented by Saad [19]. The main idea of their investigation relies on using an approximation for the centrifugal term.

Moreover, the simple conventional Nikiforov-Uvarov method [20], which received much interest, has been introduced for solving Schr$\ddot{o}$dinger equation [6], Klein-Gordon [21], Dirac [22] equations.

It is therefore the aim of this paper to present the approximate solutions of the Schr$\ddot{o}$dinger equation with the Hulth$\acute{e}$n potential in \textsl D-dimensions for $\ell\neq 0$   states using the conventional Nikiforov-Uvarov method. The paper is arranged as follows: the next section gives the Schr$\ddot{o}$dinger equation in \textsl D-dimensions; section 3 gives a brief description of the Nikiforov-Uvarov method, while the next section presents the bound-state to the  Hulth$\acute{e}$n potential. In section 5, some expectation values are calculated; and finally, the conclusions of the work are presented in section 6.\\
\pagebreak

\noindent{\bf 2.0 Schr$\ddot{o}$dinger Equation in \textsl D-dimensional Spaces.}\\

\noindent Using the \textsl D-dimensional polar cordinate with polar variable \textsl r (hyperradius) and the angular momentum variable  ${\theta_1,\theta_2,\theta_3,...,\theta_{D-2},\phi}$ (hyper angle), the Laplacian operator in the polar coordinate ${r,\theta_1,\theta_2,\theta_3,...,\theta_{D-2},\phi}$ of the $R^D$ is

$$\nabla^2_D=r^{1-D}\frac{\partial}{\partial r}\left(r^{D-1}\frac{\partial}{\partial r}\right)+ \frac{\Lambda^2_D(\Omega)}{r^2}   \eqno{(1)}$$\\ 
where $\Lambda^2_D(\Omega)$   is a partial differential operator on the unit sphere $S^{D-1}$  (Laplace-Betrami operator or the grand orbital operator) define analogously to a three-dimensional angular momentum[2]  as $\Lambda^2_D(\Omega)=-\Sigma^D_{i\geq j}(\Lambda^2_{ij})$ where \linebreak $\Lambda^2_{ij}=x_i\frac{\partial}{\partial x_j}- x_j\frac{\partial}{\partial x_i}$ for all Cartesian component $x_i$ of the \textsl \linebreak \textsl D-dimensional vector $(x_1,x_2,...,x_D)$.\\
 The \textsl D-dimensional Schr$\ddot{o}$dinger equation has the form [2,4]:
 $$-\frac{\hbar^2}{2\mu}\nabla^2_D\Psi_{n\ell m}(r,\Omega)+V(r)\Psi_{n\ell m}(r,\Omega)=E\Psi_{n\ell m}(r,\Omega)  \eqno{(2)}$$
where $\mu$ is the reduced mass and $\hbar$ is the Plank's constant.\\

\noindent {\bf 2.1 The Hyperradial Equation for the Hulth$\acute{e}$n Potential.}\\\\
The Hulth$\acute{e}$n potential is given as [7, 8, 9, 19, and 23]:
$$V_H(r)=-Z\alpha\frac{e^{-\alpha r}}{1-e^{-\alpha r}}   \eqno {(3)} $$
where $\alpha$ is the screening parameter and $Z$ is a constant which is identified with the atomic number when the potential is used for atomic phenomena.\\
Inserting Eq.(3) into Eq.(2), we have
$$r^{1-D}\frac{\partial}{\partial r}\left(r^{D-1}\frac{\partial}{\partial r}\right)\Psi_{n\ell m}(r,\Omega)+\frac{\Lambda_D^2}{r^2}\Psi_{n\ell m}(r,\Omega)+\frac{2\mu}{\hbar^2}\left[E+Z\alpha\frac{e^{-\alpha r}}{1-e^{-\alpha r}}\right]\Psi_{n\ell m}(r,\Omega)=0  \eqno{(4)}$$
Seperating the variables Eq.(4) becomes 
$$r^{1-D}\frac{d}{d r}\left(r^{D-1}\frac{d}{dr}\right)R_{n\ell}(r)-\frac{\ell(\ell+D-2)}{r^2}R_{n\ell}(r)+\frac{2\mu}{\hbar^2}\left[E+Z\alpha\frac{e^{-\alpha r}}{1-e^{-\alpha r}}\right]R_{n\ell}(r)=0  \eqno{(5)}$$
and
$$\Lambda^2_D(\Omega)Y^m_\ell(\Omega)+\beta Y^m_\ell(\Omega)=0 \eqno{(6)}$$
where the seperation constant $\beta$ is given as 
$$\beta=\ell(\ell+N+2);\hspace{0.5in} \ell=0,1,2,...  \eqno{(7)}$$
Thus, we write Eq.(5) as follow
$$R^{\prime\prime}_{n\ell}(r)+\frac{D-1}{r}R^\prime_{n\ell}(r)+ \frac{\ell(\ell+D+2)}{r^2}R_{n\ell}(r)+ \frac{2\mu}{\hbar^2}\left[E+Z\alpha\frac{e^{-\alpha r}}{1-e^{-\alpha r}}\right]R_{n\ell}(r)=0  \eqno{(8)}$$
where $R_{n\ell}(r)$  is the hyper radial part of the wave function,\textsl E is the energy eigenvalue, $\ell$ is the Orbital angular momentum quantum number. Eq.(8) is the \textsl D-dimensional hyper radial Schr$\ddot{o}$dinger equation for the Hulth$\acute{e}$n potential.\\\\
\noindent{\bf 3. The Nikiforov-Uvarov Method.}\\\\
In this section, we give a brief description of the conventional Nikiforov-Uvarov method. A more detailed description of the method can be obtained the following references [20].With an appropriate transformation $s=s(r)$,the one dimensional Schrödinger equation can be reduced to a generalized equation of hypergeometric type which can be written as follows:
$$\psi^{\prime\prime}(s)+ \frac{\tilde{\tau}(s)}{\sigma(s)}\psi^\prime(s)+ \frac{\tilde{\sigma}(s)}{\sigma^2(s)}\psi(s)=0  \eqno{(9)}$$ 
Where $\sigma(s)$and $\tilde{\sigma}(s)$ are polynomials, at most second-degree, and $\tilde{\tau}(s)$is at most a first-order polynomial. To find particular solution of Eq.(9) by separation of variables, if one deals with
$$\psi(s)=\phi(s)y_n(s),  \eqno{(10)}$$
Eq.(9)becomes
$$\sigma(s)y^{\prime\prime}_n+\tau(s)y^\prime_n +\lambda y_n =0  \eqno{(11)}$$
where
$$\sigma(s)= \pi(s)\frac{\phi(s)}{\phi^\prime(s)},  \eqno{(12)}$$,
$$\tau(s)=\tilde{\tau}(s)+2\pi(s) ,  \tau^\prime(s)<0,  \eqno{(13)}$$\\
$$\pi(s)=\frac{\sigma^\prime(s)-\tilde{\tau}(s)}{2}\pm \sqrt{\left(\frac{\sigma^\prime(s)-\tilde{\tau}(s)}{2}\right)^2-\tilde{\sigma}(s)+t\sigma(s)},  \eqno{(14)}$$
and 
$$\lambda=t+\pi^\prime(s).  \eqno{(15)}$$
The polynomial $\tau(s)$ with the parameter $s$ and prime factors show the differentials at first degree be negative. However,determination of parameter $t$ is the essential point in the calculation of $\pi(s)$. It is simply defined by setting the discriminate of the square root to zero [20]. Therefore, one gets a general quadratic equation for $t$.The values of $t$ can be used for calculation of energy eigenvalues using the following equation
$$\lambda=t+\pi^\prime(s)=-n\tau^\prime(s)-\frac{n(n-1)}{2}\sigma^{\prime\prime}(s).   \eqno{(16)}$$
Furthermore, the other part $y_n(s)$ of the wave function in Eq.(12) is the hypergeometric-type function whose polynomial solutions are given by Rodrigues relation: 
$$y_n(s)=\frac{B_n}{\rho(s)}\frac{d^n}{ds^n}[\sigma^n(s)\rho(s)]  \eqno{(17)}$$ 
where $B_n$ is a normalizing constant and the weight function $\rho(s)$ must satisfy the condition [20]
$$(\sigma\rho)^\prime =\tau\rho.   \eqno{(18)}$$\\

\noindent{\bf 4.0. Eigenvalues of the Hulth$\acute{e}$n potential in \textsl D-dimensions.}\\\\
We now want to obtain the solution to Eq.(8) using the Nikiforov-Uvarov method.If one defines
$$R_{n\ell}(r)=r^{-\frac{(D-1)}{2}}U_{n\ell}(r),    \eqno(19)$$
Eq.(8)becomes
$$U_{n\ell}^{\prime\prime}(r)+\left[\frac{2\mu}{\hbar^2}\left(E+Z\alpha\frac{e^{-\alpha r}}{1-e^{-\alpha r}}\right)-\frac{(2\ell+D-1)(2\ell+D-3)}{4r^2}\right]U_{n\ell}(r)=0. \eqno{(20)}$$

Eq.(20) is similar to the one dimensional Schr$\ddot{o}$dinger equation for the Hulth$\acute{e}$n potential, expect for the addition of the term $\frac{\ell(\ell+N-2)}{r^2}$,which is well known as the centrifugal term.To solve Eq.(20),one can consider the approximation of the centrifugal term which is valid for a small value of $\alpha $ 
$$\frac{1}{r^2}\approx\frac{\alpha^2e^{-\alpha r}}{(1-e^{-\alpha r})^2}.  \eqno{(21)}$$
With the use of Eq.(21) and the transformation $s=e^{-\alpha r}$, Eq.(20) becomes
$$U_{n\ell}^{\prime\prime}(s)+\frac{(1-s)}{s(1-s)}U_{n\ell}^\prime(s)+\frac{1}{[s(1-s)]^2}[(-\epsilon^2-\delta)s^2+(2\epsilon^2+\delta-\gamma)s-\epsilon^2]U_{n\ell}(s)=0  \eqno{(22)}$$
where
$$-\epsilon^2=\frac{2\mu E}{\alpha^2 \hbar^2},\hspace{.2in} \delta=\frac{2Z\mu}{\alpha\hbar^2},\hspace{.2in} and \hspace{.2in} \gamma=\frac{(2\ell+D-1)(2\ell+D-3)}{4}  \eqno{(23)}$$
By comparing Eqs.(9) and (22), we can define the following
$$\tilde{\tau}(s)=1-s,\hspace{.2in} \sigma(s)=s(1-s)\hspace{.2in} and \hspace{.2in}\tilde{\sigma}(s)=(-\epsilon^2-\delta)s^2+(2\epsilon^2+\delta-\gamma)s-\epsilon^2  \eqno{(24)}$$
Inserting these expressions into Eq.(14), we have
$$\pi(s)=-\frac{s}{2}\pm \frac{1}{2}\sqrt{[1+4(\epsilon^2+\delta-t)]s^2-4(2\epsilon^2+\delta-\gamma+t)s+4\epsilon^2}  \eqno{(25)}$$ 
The constant parameter $t$ can be found by the condition that the expression under the square root has a double zero, i.e. the discriminant is zero. Thus, the two possible functions for each $t$ are given as\\
$$\pi(r)=-\frac{s}{2}\pm \left\{\begin{array}{rll}
\frac {1}{2}\left [\left(2\epsilon -\sqrt{1+4\gamma}\right)s-2\epsilon\right] & \mbox{for} & t=\gamma-\delta+\epsilon\sqrt{1+4\gamma}\\\\
\frac {1}{2}\left[\left(2\epsilon +\sqrt{1+4\gamma}\right)s-2\epsilon\right] & \mbox{for} & t=\gamma-\delta-\epsilon\sqrt{1+4\gamma}\end{array}\right. \eqno{(26)}$$\\
However, for the polynomial $\tau(s)=\tilde{\tau}(s)+2\pi(s)$ to have a negative derivative, we can select the physically valid solution to be$$\pi(s)=-\frac{s}{2}-\frac {1}{2}\left[\left(2\epsilon +\sqrt{1+4\gamma}\right)s-2\epsilon\right]  \eqno{(27)}$$ for $ t=\gamma-\delta-\epsilon\sqrt{1+4\gamma}$ such that$$\tau(s)=1-2s-\left[\left(2\epsilon +\sqrt{1+4\gamma}\right)s-2\epsilon\right]. \eqno{(28)}$$
Also,by Eq.(16) we can define
$$\lambda=\gamma-\delta-\frac{1}{2}(1+2\epsilon)\left(1+\sqrt{1+4\gamma}\right)=n\left[1+2\epsilon+n+\sqrt{1+4\gamma}\right]  \eqno{(29)}$$for $n=1,2,...$
Thus, with the aid of Eqs.(23)and (29),after simple manipulations, we have the energy eigenvalue as
$$E_n=-\frac{\alpha^2\hbar^2}{2\mu}\left[\frac{1}{2}+\frac{n(n+2\ell+D-2)+(\gamma-\delta)}{2n+2\ell+D-1}\right]^2 , \eqno{(30)}$$ where $\gamma$ and $\delta$ are define in Eq.(23).We now give the following comment.\\ Comments: Eq.(30) gives the energy spectrum of the Hulth$\acute{e}$n potential in \textsl D-dimensions. However, in order to check the validity of the approximation (21), we show that Eq.(30) can be reduced  to a 3-dimensional case. That is when $D=3$ and $\gamma=\ell(\ell+1)$ and after a simple manipulation, we have
$$E_n=-\frac{\hbar^2}{2\mu}\left[\frac{(Z\mu/\hbar^2)}{n+\ell+1}-\frac{n+\ell+1}{2}\alpha\right]^2  \eqno{(31)}$$ which is in agreement with previous works [6,24]. Also, we note that if we write Eq.(30) as follows
$$E_n=-\frac{\hbar^2}{2\mu}\left[\frac{\alpha}{2}+\frac{\alpha n(n+2\ell+D-2)+(\alpha\gamma-\frac{2Z\mu}{\hbar^2})}{2n+2\ell+D-1}\right]^2,  \eqno{(32)}$$ 
and take the limit as $\alpha\rightarrow 0$, we have 
$$E_n= -\frac{\mu}{2\hbar^2}\left[\frac{2Z}{2n+2\ell+D-1}\right]^2 \eqno{(33)}$$
which is the energy eigenvalue of the Coulombic potential[29]. Evidently, this follows from the fact that $\lim_{\alpha\rightarrow 0}V_H(r)=-\frac{Z}{r}$.\\\\
\noindent{\bf 4.1. Eigenfunctions of the Hulth$\acute{e}$n potential in \textsl D-dimensions.}\\\\
In this section, we obtain the wave functions using the Nikiforov-Uvarov method. By substituting $\pi(s)$ and $\sigma(s)$ into Eq.(12), and solving the first order differential  equation
 we have
$$\phi(s)=s^\epsilon(1-s)^{v/2},\hspace{.3in}v=2\ell+D-1.  \eqno{(34)}$$
Also by Eq.(18), the weight function $\rho(s)$ can be obtained as 
$$\rho (s)=s^{2\epsilon}(1-s)^{v-1}  \eqno{(35)}$$
Substituting Eq.(35)into the Rodrigues relation (17), we have
$$y_n(s)=B_ns^{-2\epsilon}(1-s)^{1-v}\frac{d^n}{ds^n}\left[s^{n+2\epsilon}(1-s)^{n+v-1}\right]. \eqno{(36)}$$ 
Therefore, we can write the wave function $U_{n\ell}(s)$ as 
$$U_{n\ell}(s)=C_ns^\epsilon(1-s)^{v/2}P^{(2\epsilon,v-1)}_n(1-2s)  \eqno{(37)}$$
where $C_n$ is the normalization constant, and we have used the definition of the Jacobi polynomials[26],given as
$$P^{(a,b)}_n(s)=\frac{(-1)^n}{n!2^n(1-s)^a(1+s)^b}\frac{d^n}{ds^n}\left[(1-s)^{a+n}(1+s)^{b+n}\right].  \eqno{(38)}$$
To compute the normalization constant $C_n$, it is easy to show with the use of Eq.(19) that $$\int^\infty_0|R_{n\ell}(r)|^2r^{D-1}dr=\int^\infty_0|U_{n\ell}(r)|^2dr=\int^1_0|U_{n\ell}(s)|^2\frac{ds}{\alpha s}=1  \eqno{(39)}$$
where we have also used the substitution $s=e^{-\alpha r}$. Putting Eq.(37) into Eq.(39) and using the following definition of the Jacobi polynomial[26]
$$P^{(a,b)}_n(s)=\frac{\Gamma(n+a+1)}{n!\Gamma(1+a)} \ _2F_1\left(-n,a+b+n+1;1+a;\frac{1-s}{2}\right),  \eqno{(40)}$$ we arrived at 
$$C_n^2\left[\frac{\Gamma(2\epsilon+n+1)}{n!\Gamma(2\epsilon+1)}\right]^2\int_0^1s^{2\epsilon-1}(1-s)^v\ _2F_1\left(-n,2\epsilon+v+n;1+2\epsilon;s\right)ds=\alpha \eqno{(41)}$$ where $F$ is the hypergeometric function. Using the following series representation of the hypergeometric fucntion
$$_pF_q(a_1,...,a_p;c_1,...,c_q;s)=\sum_{n=0}^\infty\frac{(a_1)_n...(a_p)_n}{(c_1)_n...(c_q)_n}\frac{s^n}{n!}  \eqno{(42)}$$we have
$$C_n^2\left[\frac{\Gamma(2\epsilon+n+1)}{n!\Gamma(2\epsilon+1)}\right]^2\sum^n_{k=0}\sum^n_{j=0}\frac{(-n)_k(n+2\epsilon+v)_k}{(1+2\epsilon)_k k!}\frac{(-n)_j(n+2\epsilon+v)_j}{(1+2\epsilon)_j j!}\int_0^1s^{2\epsilon+k+j-1}(1-s)^v ds=\alpha .\eqno{(43)}$$
Hence, by the definition of the Beta function,Eq.(43)becomes
$$C_n^2\left[\frac{\Gamma(2\epsilon+n+1)}{n!\Gamma(2\epsilon+1)}\right]^2\sum^n_{k=0}\sum^n_{j=0}\frac{(-n)_k(n+2\epsilon+v)_k}{(1+2\epsilon)_k k!}\frac{(-n)_j(n+2\epsilon+v)_j}{(1+2\epsilon)_j j!} B(2\epsilon+k+j,v+1)=\alpha.  \eqno{(44)}$$
Using the relations $B(x,y)=\frac{\Gamma(x)\Gamma(y)}{\Gamma(x+y)}$ and the Pochhammer symbol \linebreak $(a)_n=\frac{\Gamma(a+n)}{\Gamma(a)}$, Eq.(44) can be written as 
$$C_n^2\left[\frac{\Gamma(2\epsilon+n+1)}{n!\Gamma(2\epsilon+1)}\right]^2\sum^n_{k=0}\frac{(-n)_k(2\epsilon)_k(n+2\epsilon+v)_k}{(1+2\epsilon+v)_k (1+2\epsilon)_k k!}\sum^n_{j=0}\frac{(-n)_j(2\epsilon+k)_j(n+2\epsilon+v)_j}{(1+2\epsilon+v+k)_j (1+2\epsilon)_j j!}=\frac{\alpha}{B(2\epsilon,v+1)}  \eqno{(45)}$$

\noindent Eq.(45) can be used to compute the normalization constants for $n=0,1,2,...$ In particular for the gound state, $n=0$, we have
$$C_0=\sqrt{\frac{\alpha}{B(2\epsilon,v+1)}}  \eqno{(46)}$$\\

\noindent{\bf 5. Some Expectation Values for the Hulth$\acute{e}$n Potential in\linebreak \textsl D-dimensons.}\\\\
We now calculate some expectation values of the Hulth$\acute{e}$n potential using the Hellmann-Feynmann theorem (HFT)[27,28].Suppose the Hamiltonian $H$ for a particular quantum system is a function of some parameters $q$, and let $E_n(q)$ and $\Psi_n(q)$ be the eigenvalues and eigenfunctions of $H(q)$ respectively, then the HFT states that$$\frac{\partial E_n(q) }{\partial q}=\langle\Psi_n(q)\arrowvert\frac{\partial H(q)}{\partial q}\arrowvert\Psi_n(q)\rangle  \eqno{(47)}$$
The effective Hamiltonian of the hyperradial function is given as
$$H=\frac{-\hbar^2}{2\mu}\frac{d^2}{dr^2}+\frac{\hbar^2}{2\mu}\frac{(2\ell+D-1)(2\ell+D-3)}{4r^2}-\frac{Z\alpha e^{-\alpha r}}{1-e^{-\alpha r}}   \eqno{(48)}$$
In order to calculate $\langle r^{-2}\rangle$, we set $q=\ell$ such that
$$\langle\Psi_n(\ell)\arrowvert\frac{\partial H(\ell)}{\partial \ell}\arrowvert\Psi_n(\ell)\rangle=\frac{\hbar^2}{2\mu}(2\ell+D-2)\langle r^{-2}\rangle  \eqno{(49)}$$ and 
$$\frac{\partial E_n}{\partial \ell}=\frac{\alpha^2\hbar^2}{8\mu}\frac{[(16\delta^2-(2n+2\ell+D-1)^4]}{(2n+2\ell+D-1)^3}  \eqno{(50)}$$  
Thus by HFT, we have
$$\langle r^{-2}\rangle=\frac{\alpha^2}{4}\frac{[(16\delta^2-(2n+2\ell+D-1)^4]}{(2\ell+D-2)(2n+2\ell+D-1)^3}  \eqno{(51)}$$
Similarly, putting $q=Z$ in Eq.(47),we obtain $\langle V(r)\rangle$ as
$$\langle V(r)\rangle=\frac{2\alpha Z}{(2n+2\ell+D-1)}\left[\frac{1}{2}+\frac{n(n+2\ell+D-2)+(\gamma-\delta)}{2n+2\ell+D-1}\right] \eqno{(52)}$$
where $\delta=2Z\mu/\alpha\hbar^2$ and $\gamma=\frac{(2\ell+D-1)(2\ell+D-3)}{4}$. Finally, we note that the expectation value $\langle T\rangle$,can also be obtained from the fact that $E_n=\langle T\rangle+\langle V\rangle$.\\

\noindent{\bf 6. Conclusions.}\\\\
In this paper,we obtained  the solutions to the \textsl D-dimensional Schr$\ddot{o}$dinger equation with the Hulth$\acute{e}$n potential within the framework of an approximation to the centrifugal term. In solving the hyperradial equation, the Nikiforov-Uvarov method was used, and the energy eigenvalues obtained was found to agree with the 3-dimensional case when $D=3$.The eigenfuctions were also obtained in terms of the Jacobi polynomials and the normalization constants were also computed in terms of the hypergeometric functions.  However,it is important to note that the approximation (21) is only valid for a small value of $\alpha $; and as $\alpha\rightarrow 0$, the results obtained approach those of the Coulombic potential.

Moreover, some expectation values for the Hulth$\acute{e}$n potential in \textsl D-dimensions were worked out using the Hellmann-Feynmann theorem. Finally, we note that the results obtained are parameter dependent.
\pagebreak

\end{document}